\documentclass[aps,pra,twocolumn,superscriptaddress]{revtex4}

\usepackage{graphicx}
\usepackage{natbib}
\usepackage[breaklinks]{hyperref}
\usepackage{amsmath}

\newcommand{\hflev}[4]{\textit{#1}$_{#2/#3}$, \textit{F}=#4}
\begin{document}


\title{Hong-Ou-Mandel interference between triggered and heralded single photons from separate atomic systems}
\author{Victor Leong}
\affiliation{Center for Quantum Technologies, 3 Science Drive 2, Singapore
  117543}
\affiliation{Department of Physics, National University of Singapore, 2 Science Drive 3, Singapore  117542}
\author{Sandoko Kosen}
\affiliation{Center for Quantum Technologies, 3 Science Drive 2, Singapore
  117543}
\affiliation{Department of Physics, National University of Singapore, 2 Science Drive 3, Singapore  117542}
\author{Bharath Srivathsan}
\author{Gurpreet Kaur Gulati}
\affiliation{Center for Quantum Technologies, 3 Science Drive 2, Singapore
  117543}
\author{Alessandro Cer\`{e}}
\affiliation{Center for Quantum Technologies, 3 Science Drive 2, Singapore  117543}
\author{Christian Kurtsiefer}
\affiliation{Center for Quantum Technologies, 3 Science Drive 2, Singapore
  117543}
\affiliation{Department of Physics, National University of Singapore, 2 Science Drive 3, Singapore  117542}
\email[]{christian.kurtsiefer@gmail.com}
\date{\today}
\begin{abstract}
We present Hong-Ou-Mandel interference of single photons generated via two different physical processes by two independent atomic systems:
scattering by a single atom, and parametric generation via four-wave mixing
in a cloud of cold atoms. 
Without any spectral filtering, we observe a visibility of $V=62\pm4\%$. 
After correcting for accidental coincidences, we obtain $V=93\pm6\%$.
The observed interference demonstrates the compatibility of the two sources,
forming the basis for an efficient quantum interface between different
physical systems.

\end{abstract}
\maketitle

\section{Introduction}
Hong-Ou-Mandel~(HOM) interference~\cite{hong:87} takes place when two
indistinguishable photons arrive simultaneously at the two inputs of a 50:50
beam splitter, making them leave together from the same output
port~\cite{Loudon:579300}. It provides a fundamental primitive for the
coherent interfacing of separate quantum systems via their emitted
photons~\cite{Briegel:1998} as an alternative to their direct
interaction~\cite{Dyckovsky:2012ea,Meyer:2015}.  It is the basis of quantum
teleportation~\cite{Bennett:1992tv,bouwmeester:97,DeMartini:2002vf} and
entanglement swapping~\cite{Zukowski:1993fs,pan:98}.

Initially developed as a sensitive tool for timing measurements, this effect
has been used for connecting separated copies of the same quantum systems with
photons:  
nonlinear crystals~\cite{Rarity:1995fz, DeRiedmatten:2003cl, Kaltenbaek:2006ch},
neutral atoms~\cite{Legero:2004vk, Beugnon:2006bj},
with a particularly high visibility between two $^{87}$Rb atoms~\cite{Hofmann:2012},
quantum dots~\cite{Santori:2002gg, Patel:2010gl},
NV centers in diamond~\cite{Bernien:2012hy},
single molecules~\cite{Kiraz:2005kv, Lettow:2010gw},
atomic ensembles~\cite{Chaneliere:2007im},
trapped ions~\cite{Maunz:2007hc},
and superconducting qubits~\cite{Lang:2013es}.
In order to observe the HOM interference,
two photons must be indistinguishable in all degrees of freedom.
The use of identical sources ensures the matching of
the temporal shape and bandwidth
of the generated photons, allowing for very high visibility when the sources are accurately synchronized. 

There are still few experimental demonstrations of HOM interference 
with single photons originating from different physical processes:
a single quantum dot and parametric down-conversion in a nonlinear
crystal~\cite{Polyakov:2011fza}, and different parametric effects in
nonlinear optical materials~\cite{McMillan:2013vm}.
These two demonstrations rely on spectral filtering in order to match
the temporal shape and the bandwidth of the generated photons.

\section{Idea}
In this work,
we demonstrate the compatibility of two single photon sources based on
$^{87}$Rb which generate single photons via two different physical processes:
scattering from a single atom~(SA) in free space, and heralding on photon
pairs prepared by parametric conversion using four-wave mixing~(FWM) in a cold
atomic vapor.

\begin{figure}
  \includegraphics[width=\columnwidth]{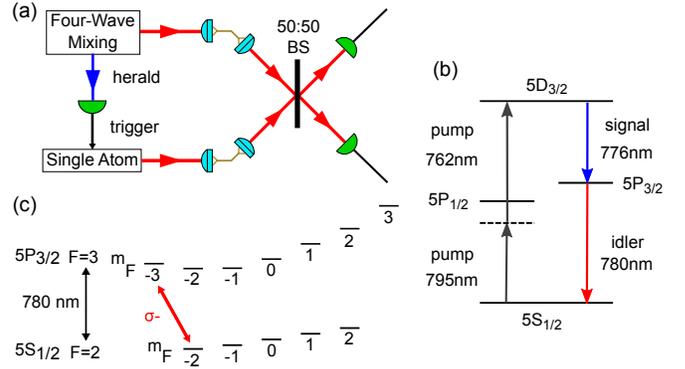}
  \caption{\label{fig:idea}(a)~Schematic representation of the Hong-Ou-Mandel
    experiment. Heralded photons from pairs generated by four-wave mixing
    in an atomic ensemble interfere with single photons generated by a
    single atom after heralding on a 50:50 beam splitter, and are detected by
    avalanche photodetectors at the 
    outputs. (b)~Simplified level scheme of the FWM process. (c)~Level scheme
    for the single atom in the dipole trap and electronic
    transition used for exciting the single atom.}
\end{figure}

As depicted in Fig.~\ref{fig:idea}(a), we combine the generated single photons on a 50:50 beam splitter. If the two photons are compatible, the HOM effect will decrease the rate of coincident events at the outputs as compared to having two completely distinguishable photons.

Both sources generate single photons with a decaying exponential temporal envelope.
For the SA source, the time constant is given by the natural linewidth of the transition~\cite{Syed:2013}, 
while for the FWM source it is 
determined by the
optical density of the atomic ensemble~\cite{Jen:2012,Srivathsan:2013}.

The timing characteristics of the two sources are determined by the generation
processes.  The FWM process generates photon pairs with Poissonian statistics,
and we obtain a heralded single photon by detecting one photon of the
pair~\cite{Clauser:1974hc,Grangier:2007ge,Hong:1986zz}, while the emission of
a single photon from the single atom is triggered by an excitation pulse. 
The detection of the heralding photons from  the FWM also serves as the
trigger for the excitation pulse of the single atom source,
effectively synchronizing the whole experiment.

\section{Experimental setup}
\begin{figure}
  \includegraphics[width=\columnwidth]{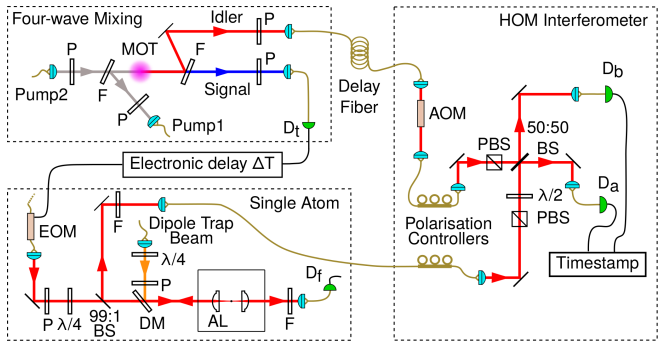}
  \caption{\label{fig:schematic}
    (Top left) Four-wave mixing setup: Pump\,1 (795\,nm) and Pump\,2 (762\,nm)  are overlapped in a copropagating geometry inside the cold cloud of  $^{87}${Rb} atoms in a Magneto-Optical Trap (MOT), generating signal (776\,nm) and idler (780\,nm) photon pairs. The detection of a signal photon heralds the presence of a single photon in idler mode, and is used to trigger the excitation of the single atom.
    (Bottom left) Single atom setup: A $^{87}${Rb} atom is trapped in
    free space between two confocal aspheric lenses (AL; numerical aperture
    0.55) with a far-off-resonant optical dipole trap
    ($\lambda=980\,\text{nm}$). After an adjustable delay time $\Delta T$
    from the trigger, an electro-optic modulator (EOM) generates an optical
    pulse to efficiently excite the single atom. The presence of an atom in the
    trap is periodically checked using APD $D_f$.
    (Right) HOM interferometer: single photons from both sources interfere at a 50:50 beam splitter (BS). An acousto-optic modulator (AOM) matches the central frequencies of both photons.
    P: polarizer, $F$: interference filters,  $\lambda$/2, $\lambda$/4: half- and quarter-wave plates, PBS: polarizing beam splitter, BS: non-polarizing beam splitter, $D_{a}$, $D_{b}$, $D_{f}$, $D_{t}$: avalanche photodetectors.
  }
\end{figure}

Figure~\ref{fig:idea}(b) shows the FWM energy level scheme: 
two pump beams at 795\,nm and 762\,nm excite the atoms 
from 5\hflev{S}{1}{2}{2}
to the 5\hflev{D}{3}{2}{3} level via a two photon transition.
The detailed experimental setup is shown in Fig.~\ref{fig:schematic}.
We select time-correlated photon pairs with wavelengths 776\,nm (signal) and 780\,nm (idler) using narrowband interference filters and collect them into single mode fibers. 
The detection of a signal photon by an avalanche photodetector (APD) $D_{t}$
heralds the presence of a single photon in the idler mode with a high
fidelity~\cite{Gulati:2014}.
The heralding efficiency of the FWM is $\approx$\,0.5\%, including all losses and the limited efficiency of the APD.

The SA source generates single photons by optically exciting the electronic transition of interest and 
collecting the consequent photon emitted by spontaneous decay~\cite{Darquie:2005fr}.
A single atom is trapped at the focus of
a far-off-resonant optical dipole trap (FORT)
obtained by focusing a Gaussian beam ($\lambda=980\,\text{nm}$) to a waist of
1\,$\mu$m using an aspheric lens (numerical aperture 0.55).
Further details of the trapping are described in~\cite{Syed:2013, Tey:2008}.
The trapped atom undergoes
molasses cooling and is optically pumped to the 5\hflev{S}{1}{2}{2}, $m_F$=-2 state.
To ensure a sufficiently long coherence time
of the prepared state,
we apply a bias magnetic field of 2\,gauss along the optical axis.
After the atom is prepared in the initial state, it can be excited to
5\hflev{P}{3}{2}{3}, $m_F$=-3 [see Fig.~\ref{fig:idea}(c)] by a short resonant optical pulse generated using a fast electro-optic modulator (EOM).
The beams used for optical pumping and excitation are collinear with the dipole trap, and are focused onto the atom by the same aspheric lens.
The excitation pulse duration $\tau_e=3$\,ns is
much shorter than the excited state lifetime $\tau_{s}=26\,\text{ns}$,
and its amplitude is set to 
maximize the excitation probability. 

The aspheric lens is also used to
collect the spontaneously emitted single photons.
The collection mode is separated from the excitation mode using a 99:1 beam splitter and is
then coupled into a single mode fiber.
The overall generation, collection and detection efficiency is
$\approx$\,0.5\%. 
We periodically check for the presence of the atom in the FORT by monitoring fluorescence with detector $D_f$; if the atom is lost, a new atom is loaded from a MOT.

The FWM setup is located in an adjacent room, approximately 15\,m away from the rest of the setup. To allow sufficient time to generate and synchronize the excitation pulse for the SA source, the heralded photon from the FWM travels through a 230\,m long fiber.

Both photons are launched into the two input ports of the HOM interferometer.
A polarizing beam splitter in each input port
transmits only horizontally polarized photons; a half-wave plate sets the relative polarizations of the photons incident on the non-polarizing 50:50 beam splitter. 
We measure a spatial mode overlap of $\approx$\,98\% between the two inputs. 
The output modes of the beam splitter are coupled into
two
single mode fibers connected to two APDs, $D_{a}$ and $D_{b}$.
\begin{figure}
  \includegraphics[width=\columnwidth]{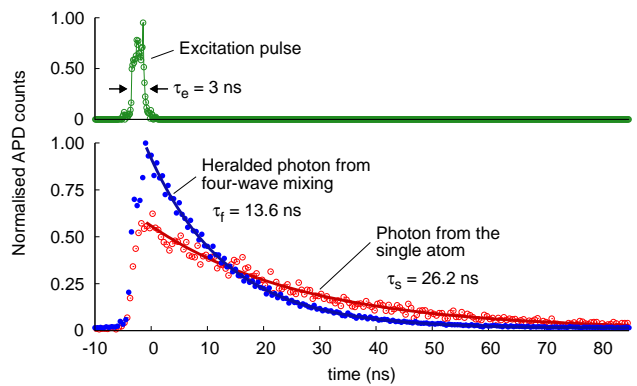}
  \caption{
    \label{fig:decay}
    (Top) Temporal profile of the excitation pulse. (Bottom) Temporal profile
    of the single photons generated by the single atom (open circles) and
    four-wave mixing (filled circles) sources. The coherence times are
    obtained from exponential fits (solid lines).}
\end{figure}

We measured the temporal envelope of the generated photons to estimate the expected visibility. 
We show these profiles in Fig.~\ref{fig:decay}, together with the temporal profile of the
pulse used to excite the single atom.
For both sources the time profile is a decaying exponential described by:
\begin{equation}\label{eq:time_envelope}
	\psi_{i}(t)\,=\,\sqrt{\frac{1}{\tau_{i}}}\,e^{-\frac{t-t_i}{2\tau_{i}}}\,\Theta(t-t_i) \textrm{ with } i=f,s\,,
\end{equation}
where $\tau_{f, s}$ are the coherence times from FWM and SA sources respectively, 
$t_s$ is the single atom excitation instant following a heralding event at $t_f$,
and $\Theta(t)$ is the Heaviside step function.
For the single atom, we confirm
$\tau_{s}=26.18\pm0.11\,\text{ns}$, corresponding to the natural linewidth of the transition.
For the FWM source,
$\tau_{f}=13.61\pm0.73$\,ns, where the uncertainty is mainly due to the drifting optical density of the atomic cloud.

In order to observe the HOM interference we also need to ensure that both photons have the same central frequency.
The single atom experiences an AC Stark shift from the dipole trap and a Zeeman shift from a bias magnetic field, 
resulting in a detuning of $\delta_{s}=76\,\text{MHz}$ from the natural transition frequency for the emitted photon.
We compensate for this detuning by shifting the central frequency of the photon coming from the FWM using 
an acousto-optic modulator (AOM).

\section{Data analysis}
\begin{figure}
  \includegraphics[width=0.9\columnwidth]{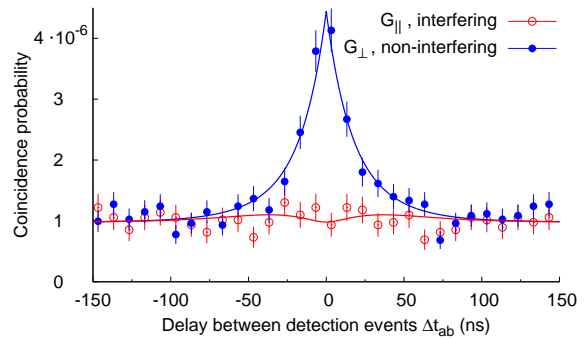}
  \caption{\label{fig:g2}
    Coincidence probability between $D_{a}$ and $D_{b}$ 
    for valid sequences measured at $\Delta T=0$. 
    The filled and open circles represent the cases where photons
    have perpendicular (non-interfering) and parallel (interfering)
    polarizations, respectively.
    The data is sorted into 10\,ns wide time bins and normalized to
    the total number of trigger events $N_t$.
    For an integration window of $T_c = -25\,\text{ns}\,\leq\,\Delta
    t_{ab}\,\leq\,25\,\text{ns}$, the interference visibility
    $V=62\pm4\%$. 
    The upper solid line represents $G_{\textrm{acc}}\,+\,A\cdot G_{\perp}(\Delta t_{ab})$ [see
    Eq.~(\ref{eq:Gperp})], and the lower solid line represents
    $G_{\textrm{acc}}\,+\,A\cdot G_{||}(\Delta t_{ab})$ [see
    Eq.~(\ref{eq:Gpll})].
    $G_{\textrm{acc}}$ is a constant offset,
    while $A$ is a scaling factor.
  }
\end{figure}

The HOM interference can be observed by comparing the probability of
coincidence $P$ between detectors $D_a$ and $D_b$ for interfering~($P_{||}$)
and non-interfering~($P_{\perp}$) photons. We adjust the relative
polarizations of the input modes from  parallel (interfering) to orthogonal
(non-interfering) by rotating a half-wave plate.
We estimate $P$ using the coincidence detection
rates. All detection events are timestamped with a temporal resolution of
125\,ps. We offset the detection times of all detectors to account for the
delays introduced by the electrical and optical delay lines, and
we only consider a detection sequence valid if either $D_{a}$ or $D_{b}$
clicks within 85\,ns of a trigger from $D_{t}$. We then sort the time delay
between detection events $\Delta t_{ab}$ into time bins of width 10\,ns
and normalize the distribution by dividing by 	the total number of trigger
events 	$N_t$ over the measurement time:
\begin{equation}
  G(\Delta t_{ab})\,=\,\frac{N_{ab|t}(\Delta t_{ab})}{N_t}\,.
\end{equation}
The measured $G_{\perp}$ and $G_{||}$ are shown in Fig.~\ref{fig:g2}.
For $|\Delta t_{ab}|\lesssim50\,$ns, the coincidence probability for 
non-interfering photons increases significantly above the background 
at large $|\Delta t_{ab}|$, while it remains at an almost constant level for the
interfering case. To
quantify this observation, we define a visibility $V$ for the HOM
interferometer as:
\begin{equation}\label{eq:vis}
  V=1-P_{||}/P_{\perp},
\end{equation}
where the probabilities $P$ are obtained
by a sum over the time bins within a coincidence window $T_c$:
\begin{equation}\label{eq:norm_P_discrete}
  V\,=\,1-\frac{\sum\limits_{T_c}G_{||}(\Delta t_{ab})}
  {\sum\limits_{T_c}G_{\perp}(\Delta t_{ab})}\,.
\end{equation}
The choice of $T_c$ determines
the influence of the accidental count rates on the visibility.
Similar to what has been used in the past~\cite{Maunz:2007hc}, 
we choose $T_c=-25\,\text{ns}\leq\Delta t_{ab}\leq 25\,\text{ns}$,
a window long enough to include the longer of the two photon coherence times, 
resulting in $V=62\pm4\%$.
\section{Theory - Time envelope matching}
The probability of coincidence events for unit time $G(\Delta t_{ab})$
in the non-interfering case, i.e., photons with orthogonal  polarization,
is given by adding probabilities for independent pair events:
\begin{align}\label{eq:Gperp}
  G_{\perp}(\Delta t_{ab})\,=\,\frac{1}{4}\int\limits^{\infty}_{-\infty} &|\psi_{f}(t)\,\psi_{s}(t+\Delta t_{ab})|^2 \nonumber \\ 
  & + |\psi_{f}(t+\Delta t_{ab})\,\psi_{s}(t)|^2\,dt \,.
\end{align} 
When the two incident photons have identical polarizations, their pair amplitudes interfere (with the minus sign determined by one of the reflections on the beam splitter):
\begin{align}\label{eq:Gpll}
  G_{||}(\Delta t_{ab})\,=\,\frac{1}{4}\int\limits^{\infty}_{-\infty} 
  &|\psi_{f}(t) \psi_{s}(t+\Delta t_{ab})\nonumber \\
  &-\,\psi_{f}(t+\Delta t_{ab}) \psi_{s}(t)|^2\,dt\,.
\end{align}
The total probability $P$
is obtained by integrating over time:
$P=\int G(\Delta t_{ab})d(\Delta t_{ab})$.
In the non-interfering case, as expected,
we obtain $P_{\perp}=\frac{1}{2}$.
In the interfering case,
for $\Delta T\,=\,0$, i.e., when the heralding time and the single atom excitation are synchronized,
$P_{||}=\frac{(\tau_s-\tau_f)^2}{2(\tau_s+\tau_f)^2}$.
Using these results, Eq.~(\ref{eq:vis}) reduces to:
\begin{equation}
  V=\frac{4\tau_s\tau_f}{(\tau_s+\tau_f)^2}.
\end{equation}
Using the measured values for $\tau_s$ and $\tau_f$,
we obtain an expected visibility of $90.0\pm1.5\%$.
To properly compare it with the one measured experimentally, we 
choose a large integration window $T_c = -75\,\text{ns}\,\leq\,\Delta t_{ab}\,\leq\,75\,\text{ns}$
and correct for accidental coincidences $G_{\textrm{acc}}$.
We obtain a corrected visibility of $V=93\pm6\%$, which is compatible with the
expected value.
\section{HOM dip}
\begin{figure}
  \includegraphics[width=0.9\columnwidth]{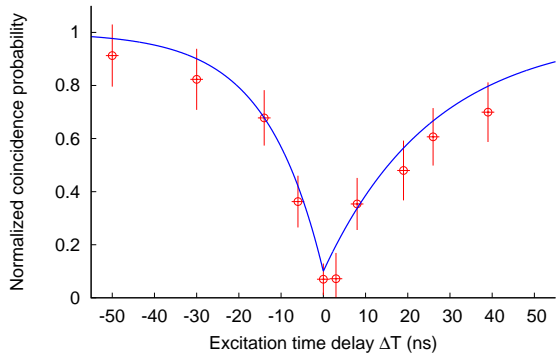}
  \caption{\label{fig:hom_dip}Normalized coincidence probability $P_{||}/P_{\perp} = 1-V$, corrected for accidental coincidences, showing the ``HOM dip''.
    The solid line shows expected values obtained from Eq.~(\ref{eq:dip}).
  }
\end{figure}
We can also vary the degree of interference by changing the delay $\Delta T$ between the heralding time $t_f$ and the single atom excitation time $t_s$.
To maintain a constant rate of
two photon events as we vary $\Delta T$,
$T_c$ has to be much larger than maximum value of $|\Delta T|$ used in the experiment.
As before, we choose $T_c=150\,\text{ns}$ and
subtract $G_{\textrm{acc}}$ from the measured $G_{\perp}$ and $G_{||}$.
  
In Fig.~\ref{fig:hom_dip} we plot the ratio $P_{||}/P_\perp$, and
observe the familiar HOM dip~\cite{hong:87}.
From Eqs.~(\ref{eq:Gperp}) and~(\ref{eq:Gpll}) we can derive the shape of the dip:
  \begin{equation}\label{eq:dip}
    \frac{P_{||}}{P_\perp} = 1-\frac{4\tau_s\tau_f\,e^{\Delta T/\tau}}{(\tau_s+\tau_f)^2}    
    \text{ with }
      \begin{cases}
          \tau = -\tau_s & \text{if } \Delta T \geq 0 \\
          \tau = \tau_f & \text{if } \Delta T < 0.
        \end{cases}
  \end{equation}
The dip is slightly asymmetric due to the different coherence times $\tau_f,
\tau_s$ in the asymmetric photon profiles in Eq.~(\ref{eq:time_envelope}).
Using Eq.~(\ref{eq:dip}) and the measured values for $\tau_f$ and $\tau_s$,
we obtain the solid line plotted in Fig.~\ref{fig:hom_dip}.
Most of the
measured points lie within one standard deviation of this line.

\section{Conclusion}
In conclusion, we have observed HOM interference between a triggered single photon source based on a single $^{87}$Rb atom, and a heralded single photon source based on four-wave mixing in a cold $^{87}$Rb cloud.

These two sources, though based on the same atomic species,
generate quantum light through two different processes.
Without any spectral filtering, we observe a HOM visibility of $V=62\pm 4\%$.
Correcting for accidental coincidences due to the limited collection efficiencies of the two sources,
the measured visibility is $93\pm6\%$, a value compatible with the expected $90.0\pm1.5\%$.

The observed interference demonstrates the compatibility of the spectral and
timing characteristics of our two sources. 
This is a fundamental requisite for the
transfer of quantum information between the two, and ultimately for the
realization of quantum networks to generate entanglement between separated
nodes~\cite{kimble:2008}
made up of different physical systems.

\begin{acknowledgments}
We acknowledge the support of this work by the National Research Foundation
(partly under grant No. NRF-CRP12-2013-03) \& Ministry of Education in
Singapore.
\end{acknowledgments}

\bibliographystyle{apsrev4-1}
\bibliography{hom.bib}{}
\end{document}